\documentclass[nofootinbib,tightenlines,aps,twocolumn,superscriptaddress,prl,longbibliography]{revtex4-1}

\usepackage[utf8]{inputenc}
\usepackage[usenames,dvipsnames]{color}
\usepackage{amssymb,amsmath}
\usepackage{framed}
\usepackage{float}
\usepackage{array}
\usepackage[normalem]{ulem}
\usepackage{xspace}
\definecolor{DarkBlue}{rgb}{0.1,0.1,0.5}
\usepackage[colorlinks=true,breaklinks, linkcolor= DarkBlue,citecolor= DarkBlue,urlcolor= DarkBlue]{hyperref}

\usepackage{mathtools}
\usepackage{tikz}
\usetikzlibrary{calc}
\usetikzlibrary{arrows,shapes}

\renewcommand{\H}{\mathcal{H}}
\newcommand{\tH}{\widetilde{\H}}
\newcommand{\U}{\mathcalus{U}}
\newcommand{\B}{\mathcalus{B}}
\newcommand{\I}{\imath}
\newcommand{\J}{\,\I^\dagger}

\newcommand{\ket}[1]{|{#1} \rangle}

\newcommand{\fet}[1]{| {#1} \rangle^{\!\text{\tiny F}}}

\newcommand{\anc}{\mathrm{anc}}

\newcommand{\calu}{\mathcalus{U}}

\renewcommand{\imath}{\mathcalus{I}}
\renewcommand{\phi}{\varphi}

\usepackage{mathptmx}
 \makeatletter
 \def\@viiipt{8.5}
 \def\@ixpt{9.5}
 \def\@xpt{10.5}
 \def\@xipt{11.5}
 \def\normalsize{\@setfontsize\normalsize\@xpt{12}}
 \def\small{\@setfontsize\small\@ixpt{11}}
 \def\footnotesize{\@setfontsize\footnotesize\@viiipt{9.5pt}}%
 \def\large{\@setfontsize\large{12.5}{14pt}}%
 \makeatother

\DeclareSymbolFont{cmsy}{OMS}{cmsy}{m}{n}
\DeclareSymbolFontAlphabet{\mathcalus}{cmsy}
\DeclareSymbolFontAlphabet{\mathcalus}{cmsy}
\usepackage[scaled=0.88]{helvet}

\DeclareFontFamily{U} {MnSymbolA}{}
\DeclareFontShape{U}{MnSymbolA}{m}{n}{
  <-6> MnSymbolA5
  <6-7> MnSymbolA6
  <7-8> MnSymbolA7
  <8-9> MnSymbolA8
  <9-10> MnSymbolA9
  <10-12> MnSymbolA10
  <12-> MnSymbolA12}{}
\DeclareFontShape{U}{MnSymbolA}{b}{n}{
  <-6> MnSymbolA-Bold5
  <6-7> MnSymbolA-Bold6
  <7-8> MnSymbolA-Bold7
  <8-9> MnSymbolA-Bold8
  <9-10> MnSymbolA-Bold9
  <10-12> MnSymbolA-Bold10
  <12-> MnSymbolA-Bold12}{}

\DeclareSymbolFont{MnSyA} {U} {MnSymbolA}{m}{n}

\DeclareMathSymbol{\ewarrow}{\mathrel}{MnSyA}{16}
\DeclareMathSymbol{\nsarrow}{\mathrel}{MnSyA}{17}
\DeclareMathSymbol{\neswarrow}{\mathrel}{MnSyA}{18}
\DeclareMathSymbol{\nwsearrow}{\mathrel}{MnSyA}{19}

\renewcommand{\updownarrow}{\nsarrow}
\renewcommand{\leftrightarrow}{\ewarrow}
\newcommand{\zp}{\leftrightarrow}
\newcommand{\zd}{\neswarrow}
\newcommand{\od}{\nwsearrow}
\newcommand{\op}{\mathclap{\quad\updownarrow}\phantom{\leftrightarrow}}   %

\usepackage{theorem}
\newtheorem{theorem}{Theorem}

\theoremstyle{definition}

\newcommand\pulse[2]{%
		\draw #1
		sin +(0.08,0.2*#2) cos +(0.08,-0.2*#2)
		sin +(0.15, #2) cos +(0.15,-#2)
		sin +(0.08,0.2* #2) cos +(0.08,-0.2* #2)}

\newcommand\rotatevacuum[2]{%
		\begin{scope}[densely dotted,thick,rotate around={-90:#1}] \pulse{#1}{#2}; \end{scope}}

\newcommand\rotatepulse[2]{%
		\begin{scope}[rotate around={-90:#1}]
		\draw #1
			sin +(0.08,0.2*#2) cos +(0.08,-0.2*#2)
			sin +(0.15, #2) cos +(0.15,-#2)
			sin +(0.08,0.2* #2) cos +(0.08,-0.2* #2);
		\end{scope}}
		
\newcommand\inter{%
	\draw [thick](-0.5,0) -- (9,0);
	\draw[thick] (2,0) -- (2,2) -- (6,2) -- (6,0);  %
	\draw[thick,dashed] (2,0) -- (2,-2);
	\draw[thick] (6,0) -- (6,-3);
	\draw[line width=2.25pt,black!50] 
		(1.75,-0.25)  --  node (bs1) {} (2.25,0.25) 
		(5.75,0.25)-- node (bs2) {} (6.25,-0.25);
	\node (phase) at (4,2) [circle,fill=black!40,draw=black,thick] {};
	\draw[line width=2pt]
		(1.75,1.75)-- node (mir1) {} (2.25,2.25)
		(5.75,2.25)-- node (mir2) {} (6.25,1.75);	}

\newcommand\papl[1]{\fbox{\ensuremath{\vphantom{o}\smash{#1}}}}

\newcommand\nsection[1]{\smallskip\textbf{\textit{#1.}}}

\theoremheaderfont{\bfseries\itshape}

\begin{document}

\abovedisplayskip=0.5ex plus 0.2ex minus 0.2ex
\belowdisplayskip=0.5ex plus 0.2ex minus 0.2ex
\setlength\theorempreskipamount{0.7ex plus 0.3ex minus 0.2ex}
\setlength\theorempostskipamount{0.7ex plus 0.3ex minus 0.2ex}

\title{Attacks on Fixed Apparatus Quantum Key Distribution Schemes}

\author{Michel Boyer}
\affiliation{ D\'epartement IRO, Universit\'e de Montr\'eal (Qu\'ebec), Canada 
 \email{Michel.Boyer@umontreal.ca}
}
\author{Ran Gelles}
\affiliation{Computer Science Department, University of California, Los Angeles, USA 
 \email{gelles@cs.ucla.edu}
}
\author{Tal Mor}
\affiliation{ Computer Science Department, Technion, Haifa, Israel 
 \email{talmo@cs.technion.ac.il}
}

\setcounter{footnote}{0}   %


\begin{abstract}

We consider quantum key distribution 
implementations in which the receiver's apparatus 
is fixed and does not depend on \emph{his} 
choice of basis at each qubit transmission.
We show that, although theoretical quantum key distribution 
is proven secure, such implementations are totally insecure against 
a strong eavesdropper that has one-time (single) access to the 
receiver's equipment.  The attack we present here, 
the ``fixed-apparatus attack''  causes a potential risk to the 
usefulness of several recent implementations.

\end{abstract}

\maketitle


\nsection{I. Introduction}
Quantum key distribution (QKD) is probably the best known application of quantum cryptography,
for it has already given rise to commercial implementations for securing communications.
In most QKD protocols a sender (Alice), 
prepares non orthogonal quantum states
to be measured by the receiver (Bob). The security of the transmission is
in principle guaranteed by the fact that an eavesdropper cannot spy the state being sent without
inducing errors and be detected. Such protocols comprise the BB84 protocol~\cite{BB84},
the B92 protocol~\cite{B92}, a six-state protocol~\cite{Brus98}, etc.

Given a mathematical description of the BB84 protocol, 
its security can indeed be proven rigorously~\cite{LC99,SP00,Mayers01,BBBMR06,BHLMO05,RK05}. Its physical
implementations may nevertheless be insecure;
for instance if states 
are encoded as photon pulses, 
critical security problems emerge 
from pulses containing two photons~\cite{BLMS00}.

It was generally taken for granted that if
sources whose states are all ideal qubits
in~$\mathcal{H}_2$ could be guaranteed,
so would be the security of BB84 implementations.
Researchers then understood that this is not the case,
the problem now is lying on the receiver's side.
Let $\mathcal{H}_A$~be the Hilbert space corresponding to those 
states~$\ket{\phi}$ 
Alice sends to Bob according to their protocol.
The eavesdropper (Eve) may send Bob states 
in a Hilbert space larger than~$\mathcal{H}_A$ and by doing so, 
alter the assumed behavior of his device.
See the fake-state attack~\cite{HM05},  
the Trojan pony attack~\cite{GLLP04,HLP08},
and the reversed-space attack~\cite{QSA07,GM12}.
This problem is inherent when using photons, 
since even if Alice sends an ideal qubit, 
each pulse can potentially contain less than or more than 
a single photon and be shifted in time (space, frequency), 
hence her ideal space~$\mathcal{H}_2$ 
is merely a subspace of a larger Hilbert space %
potentially exploitable by Eve.

\nsection{II. The Fixed-Apparatus Attack}\label{sec:fattack}%
\footnote{A preliminary version of this work appeared in \cite{BGM12}.}
Protocols (in particular BB84) require Bob to make random choices. 
In BB84
this choice is between measuring in  
the computation basis or in the
Hadamard basis. 
This can also be described as randomly choosing a unitary transformation 
(either the identity %
or the Hadamard transformation) followed
by a \emph{fixed} measurement  in the computation basis. 
That random choice is Bob's input to the protocol. 
In most automated QKD implementations, a random number generator or a pseudo
random number generator is used by Bob for generating his random input. 
However, to guarantee a faster bit-rate in some implementations, 
those random choices are made by the measurement itself. 
Indeed,  apparata for which there is no (random) input 
from Bob have been proposed and implemented
in the literature~\cite{ZBGR98,BHKLLMNPS98,HNMP02,HNDP02,RTGK02,KZHWGTR02,ATMYRBBPG04,WASST03,NHN03,NHN04,JS06}. 
Such an apparatus, with no (random) input from Bob, 
is what we call a ``fixed apparatus'';
it simply gives outputs. 

A simple behavior of Bob's fixed-apparatus procedure 
can be described as follows.
When Alice sends some state $\ket{\phi}\in \mathcal{H}_A$, Bob adds an ancillary state $\ket{0}_\mathrm{anc}\in \mathcal{H}_{\anc}$ and performs a measurement, in the computation basis, of~$\U_B \ket{\phi}_A\ket{0}_{\anc}$, where $\mathcal{H}_B=\mathcal{H}_A\otimes\mathcal{H}_\text{anc}$ and $\U_B:\mathcal{H}_B\to \mathcal{H}_B$ is a unitary. 
If Eve has secret access to~$\mathcal{H}_\text{anc}$, 
nothing prevents her from generating a state in~$\mathcal{H}_B$ 
instead of sending a state in~$\mathcal{H}_A$, 
and Bob may be unable to notice or prevent it due to the 
fact that his apparatus is fixed.
Then,
each state $\ket{r} \in \mathcal{H}_B$ measured by Bob 
can be ``reversed in time'' to yield a specific state 
{$\U_B^\dag\ket{r}$}
in $\mathcal{H}_A\otimes\mathcal{H}_\text{anc}$ that, if generated by Eve 
and given to Bob as input, results in Bob measuring exactly the state~$\ket{r}$. 
Hence, once Eve controls $\mathcal{H}_\text{anc}$ and 
$\mathcal{H}_A$, she has full control on Bob's measured states.

This type of an attack is a special case of the reversed-space 
attack~\cite{GM12,QSA07},
in which Eve performs her attack on a larger space,
obtained by using the time reversal symmetry of quantum theory~\cite{ABL64,AV90}.
Note that Eve only sends information into Bob's lab but is not 
getting information out of it, thus it is also a special type of a 
Trojan Pony attack~\cite{GLLP04,HLP08}. 

In a more realistic case which is particularly relevant when the quantum 
carriers are photons, %
the space~$\mathcal{H}_A$ 
spanned by Alice's states is a subspace of a larger space~$\tH_A$, 
which also contains basis states corresponding to a vacuum pulse 
(0 photons) or pulses with two or more 
photons, or pulses sent shifted in time, etc.
Similarly, we can consider a photonic space~$\tH_{\anc}$ 
that includes the ancillary space added by 
Bob. 
The choice of the two enclosing spaces $\tH_A$ and $\tH_{\anc}$ 
is somewhat arbitrary (namely Eve has a lot of freedom in defining them), 
as long as they are large enough so that
$\tH_A \otimes \tH_{\anc}$ 
contains $\U_B^\dag \ket{r}$ 
for any state $\ket{r}$ measured by Bob.

Such  ``inclusions'' of spaces within larger spaces need merely be \emph{unitary embeddings} (\emph{isometries}) i.e.\@  linear
maps preserving inner products;  
thus, we assume the ``inclusion'' $\I_A:\H_A\to \tH_A$ is an isometry. Also
$\I_\mathrm{anc}:\tH_A \to \tH_A\otimes\tH_\mathrm{anc}$ defined by $\I_\mathrm{anc}\ket{\phi} = \ket{\phi}\ket{0}_\mathrm{anc}$ 
and corresponding to attaching an ancilla is a unitary embedding: 
it is linear and preserves inner products. 
When  $\imath:\H_1\to\H_2$ is a unitary embedding,  the subspace~$\imath(\mathcal{H}_1)$ 
of~$\mathcal{H}_2$ spanned by all $\imath\ket{\phi}$ for $\ket{\phi}$ in $\H_1$ is such
that $\I\I^\dagger\ket{\psi} = \ket{\psi}$ for all states $\ket{\psi}$ in $\I(\H_1)$. %

Let us summarize the process (see Table~\ref{table:spaces}):
Alice sends states from $\ket{\phi}\in\mathcal{H}_A$ to Bob,
who  obtains $\I\ket{\phi}$
where $\I:\mathcal{H}_A\to \tH_A\otimes \tH_\text{anc}$ 
is the embedding %
defined by $\I\ket{\phi}=\I_\mathrm{anc}\I_A\ket{\phi}$.
Bob processes $\I\ket{\phi}$ 
using a unitary map $\U_B:\tH_A\otimes \tH_\text{anc}\to \tH_A\otimes \tH_\text{anc}$,
and then he performs a complete measurement of
a space $\mathcal{H}_B \subseteq \tH_A\otimes \tH_\text{anc}$  in a publicly known
basis~$\mathcalus{B}$.
Quite often, for simplicity, we only care about the operation of $\U_B$ 
on a subspace
$\tH$ of $\tH_A \otimes \tH_\text{anc}$ such that
$\I(\H_A)\subseteq \tH$ ($\H_A$ is embedded in $\tH$) and
$\H_B^R \equiv \U^\dagger(\H_B) \subseteq \tH$ 
(all ``reversed states'' from~$\H_B$ are in~$\tH$).

\begin{table}[H]
\begin{tabular}{p{1.2cm}l} 
\textbf{\textit{Space}} & \textbf{\textit{Meaning}} \\
\hline 
$\H_A$ & The span of Alice's States \\
$\H_B$ & The space \emph{measured} by Bob \\
$\widetilde {\H}_{A}$ & $^*$A ``realistic'' space that includes~$\H_A$ \\
$\tH_{\anc}$ & $^*$An ancillary space added by Bob \\
$\tH$ & The space on which Bob's transformation %
is relevant\\
$\H_B^R$ & %
The ``reversal'' of Bob's measured space \\
\hline
\end{tabular}\\

{\footnotesize
\noindent$^*$ There is a degree of freedom in choosing $\tH$ and
$\tH_{A}$ and $\tH_{\anc}$ under the constraint that 
 $\H_A$ is embedded in $\tH$  and %
$\H_B^R \subseteq \tH \subseteq \tH_{A}\otimes \tH_{\anc}$.}
\caption{The Various Relevant Spaces}
\label{table:spaces}
\end{table}
\smallskip

If Eve knows about $\tH_A$ (which is always prone to her attack)
and can access $\tH_\text{anc}$ then nothing prevents her from generating 
any state of her choice in
$\tH_A\otimes\tH_\text{anc}$ instead of sending a state in~$\mathcal{H}_A$.
This leads to our attack.
\begin{theorem}[fixed-apparatus attack]\label{thm:main}
If for all basis states $\ket{r}$ of a given basis $\mathcalus{B}$ of $\mathcal{H}_B$, 
Eve can produce a state $\ket{r}^R  \in \H_B^R \equiv \U_B^\dag(\H_B)$ 
such that Bob processes and measures $\U_B\ket{r}^R= \ket{r}$, 
then she can get full information of Bob's measured outcomes without changing their statistics.
\end{theorem}
\textit{Proof.}
Note that 
Eve knows the operation of $\U_B$ (due to knowing Bob's setup).
Eve captures the incoming state~$\ket{\varphi}$ and applies 
the same measurement as Bob on~$\U_B\imath\ket{\varphi}$, 
getting $\ket{r}\in\B$.  She then sends Bob the state~$\ket{r}^R$. Bob applies
$\U_B$ to $\ket{r}^R$  which produces the state $\ket{r}$ on which 
Bob's complete measurement in the basis $\B$  yields~$\ket{r}$, the same value obtained by Eve.
\hfill$\square$

Note that even though Eve's measurement outcome is not always 
identical to what Alice sent, 
Eve's information is always identical to Bob's,
and whenever Alice and
Bob have a matching basis, all of them hold the same bit value.

\nsection{Application: an attack on a polarization-based scheme}
Polarization-based QKD is realized by many experiments (e.g.~\cite{MBG93,MZG96,zbinden98,GRTZ02} and references therein), 
some of which~\cite{BHKLLMNPS98,HNMP02,HNDP02,RTGK02,KZHWGTR02,ATMYRBBPG04}
suffer from the above weakness.
Specifically, we now analyze an implementation of a 
polarization-based BB84, 
based on polarization beamsplitters (PBS). 
Each qubit sent by Alice either has  a horizontal/vertical polarization 
(the {$+$} %
basis), or  a diagonal polarization, i.e., $\pm45^\circ$ (the  {{$\times$}} %
basis).
In the standard implementation Bob randomly choses whether to measure
each arriving qubit using the {$+$} basis setup, or the rotated setup.
The measurement itself is implemented using two
PBSs: one separates horizontal polarization from a vertical one and the other
separates $+45^\circ$ from $-45^\circ$ (say, via a polarization rotator that
rotates the photon's polarization by $45^\circ$).

\begin{figure}[htb]
\begin{center}
\includegraphics[scale=1]{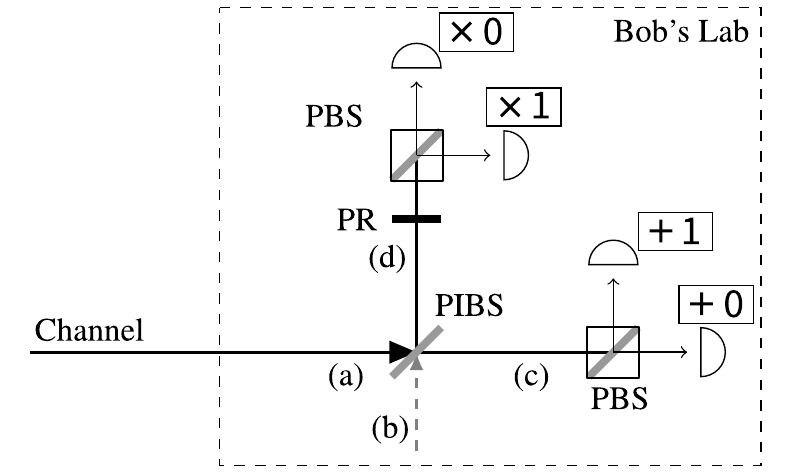}
\end{center}
\caption{%
A fixed apparatus for polarization-based BB84.
 PIBS=polarization independent beam splitter, PBS=polarization beam splitter, PR=polarization rotator.}
\label{fig:multiplex}
\end{figure}
In a \emph{fixed-apparatus} implementation,
Bob's apparatus (outlined in Fig.~\ref{fig:multiplex}) is such that both PBSs are connected to the
channel via a standard (polarization independent) beamsplitter
(PIBS), with one blocked
input arm~\cite{BHKLLMNPS98,HNMP02,HNDP02,RTGK02,KZHWGTR02,ATMYRBBPG04}.
In this case, Alice's qubit  $\ket{\phi}\in \H_A=\H_2$ is sent over the channel (arm~$a$)  
while a vacuum ancilla is implicitly added by the blocked arm of the PIBS (arm~$b$).
The embedding of 
$\mathcal{H}_A$ into the (a--b) inputs system of the PIBS is given by
$\mathcalus{I}\ket{\phi} = \ket{\phi}_a\ket{V}_b$,
where $\ket{V}$ denotes an empty (vacuum) pulse. 

We arbitrarily choose a specific embedding into 9-dimensional space, 
$\I:\mathcal{H}_A\to \tH_A\otimes \tH_\text{anc}$, 
and we shall soon see that it is the 
minimal possible embedding containing $\H_B$: 
$\tH_A$ 
is spanned by $\{\ket{V}_a,\ket{\zp}_a,\ket{\op}_a \} $
corresponding to a  vacuum state, horizontal and vertical polarization, respectively.
Similarly
$\tH_\anc$ is the span of the basis states $\{\ket{V}_b,\ket{\zp}_b,\ket{\op}_b \}$.

Bob's transformation $\U_B$ in this simple example is merely the PIBS, transferring the {(a--b)} input system into the {(c--d)} output system, 
along with the PR, and is well defined on $\tH_A\otimes \tH_\text{anc}$. 
We only need to care about 
pulses with at most a single photon,
thus (for $\ket{\phi}$ a polarization qubit) we can let $\tH = \{
\ket{\phi}_a\ket{V}_b , 
\ket{V}_a\ket{\phi}_b , \ket{V}_a\ket{V}_b
\}$, 
a five dimensional subspace of the 9-dimensional space %
$\tH_A\otimes\tH_\mathrm{anc}$.
The transformation for states in~$\tH$ is given by
\begin{equation}
\begin{split}\label{horiz}
\U_B \ket{\phi}_a\ket{V}_b &= \frac{1}{\sqrt{2}}\Big[\phantom{\mathrm{i}}\ket{\phi}_c\ket{V}_d + \mathrm{i}\ket{V}_c\mathrm{PR}\ket{\phi}_d\Big]  \\
\U_B\ket{V}_a\ket{\phi}_b &= \frac{1}{\sqrt{2}}\Big[\mathrm{i}\ket{\phi}_c\ket{V}_d + \phantom{\mathrm{i}}\ket{V}_c\mathrm{PR}\ket{\phi}_d\Big] 
\end{split}
\end{equation}
as well as $\U_B \ket{V}_a\ket{V}_b =  \ket{V}_c\ket{V}_d$. 
We could furthermore remove the $\ket{V}_a\ket{V}_b$ state from $\tH$ with no influence on the result.
The polarization rotation is given by
\(
\mathrm{PR}\ket{{\zd}} = \ket{\zp} 
\)
and 
\(
\mathrm{PR}\ket{\od} = e^{\mathrm{i}\theta}\ket{\op} 
\).

Bob's space~$\H_B$ is the four-dimensional space defined by 
the four orthogonal states measured by the four detectors.
Using our knowledge of $\U_B$, we can ``reverse'' each one of the four 
basis states $\ket{r}\in \H_B$ measured by Bob, and obtain
$\ket{r}^R \in \tH$. %
Bob's reversed space $\H_B^R$
is the space spanned by the four states $\ket{r}^R$ 
presented in Table~\ref{table:polattack}. 
See Section~\hyperref[app:BS]{A} in the Supplemental Material for full details.
\begin{table}[h]
\begin{center}
\begin{tabular}{llc}
\ $m$ & $\phantom{e^{\mathrm{i}\theta}}\ket{r}$ & $\ket{r}^R = \U_B^\dag\ket{r}$  
\\ \hline\hline
$(+,0)$\quad\quad & 
  $\phantom{e^{\mathrm{i}\theta}}\ket{\zp}_c\ket{V}_d\qquad$ &
  $(\phantom{-\mathrm{i}}\ket{\zp}_{a}\ket{V}_{b} -\mathrm{i}\ket{V}_{a}\ket{\zp}_{b}) / \sqrt{2}$ \\
$(+,1)$ & 
  $\phantom{e^{\mathrm{i}\theta}}\ket{\op}_c\ket{V}_d$ & 
  $(\phantom{-\mathrm{i}}\ket{\op}_{a}\ket{V}_{b} -\mathrm{i}\ket{V}_{a}\ket{\op}_{b}) / \sqrt{2} $\\
$(\times,0)$ & 
 $\phantom{e^{\mathrm{i}\theta}}\ket{V}_c\ket{\zp}_d$ & 
 $(-\mathrm{i}\ket{\zd}_{a}\ket{V}_{b} +\phantom{\mathrm{i}}\ket{V}_{a}\ket{\zd}_{b}) / \sqrt{2} $  \\
$(\times, 1)$ & 
 $e^{\mathrm{i}\theta}\ket{V}_c\ket{\op}_d$ & $(-\mathrm{i}\ket{\od}_{a}\ket{V}_{b} +\phantom{\mathrm{i}}\ket{V}_{a}\ket{\od}_{b}) / \sqrt{2} $  \\
\end{tabular}

\end{center}
\caption{Bob's measured states and their reversal}
\label{table:polattack}
\end{table}

In order to have full control on Bob's space $\H_B$, all Eve
has to do is to access arm $b$ of the beamsplitter (that is supposed to be blocked). 
By generating $\ket{r}^R$ not only does Eve \emph{choose}
which arm the photon goes to (arm $c$ or arm $d$) but also which detector  clicks.

The above attack can be generalized to any QKD or quantum cryptography
setting in which a beamsplitter 
is used for replacing Bob's true choice of basis.
Note that in the above example, we could have chosen $\tH=\H_B^R=\H_B$.
This is not true in general.

\nsection{III. The Fixed-Apparatus Attack on a non-trivial reversed space}\label{sec:nontrivial}
In a more general case,
for instance when using
Mach-Zehnder interferometer as we show below,
the operation applied by Bob
(after the embeddings) is not  described by a unitary
operator on $\tH_A\otimes\tH_\mathrm{anc}$.              
Bob's apparatus however always acts linearly on input
states and preserves inner products: it thus gives rise
to an isometry (unitary embedding).
In such cases, an extension of 
Theorem~\ref{thm:main} is needed.

Let us denote $\I_B$ this embedding replacing~$\U_B$; it is defined on some space $\tH \subseteq \tH_A\otimes \tH_\text{anc}$, and it can be deduced from Bob's setup.
Because~$\I_B$ is fully defined by Bob's apparatus (similarly to $\U_B$ in the earlier
section),
it is known to Eve.
We (and Eve) choose 
$\tH$ %
to ``contain''~$\H_A$, to ``contain'' relevant
vacuum ancilas that Bob adds, to be minimal, and yet to be such that 
$\H_B \subseteq 
\I_B(\tH)$
which implies that for each state $\ket{r}$ measured by Bob,
there is unique state $\ket{r}^R$ in 
$\tH$ %
(namely $\ket{r}^R = \I_B^\dagger\ket{r}$)
such that 
$\I_B (\ket{r}^R) = \ket{r}$.
As before, the span of all states $\ket{r}$ is  $\H_B$.
The span of all states $\ket{r}^R$ is now $\H_B^R = \I_B^\dagger(\H_B)$.
 
\begin{theorem}[generalized fixed-apparatus attack]\label{thm:gen1}
If for all basis states $\ket{r}$ of a given basis $\mathcalus{B}$ of $\mathcal{H}_B$, 
Eve can produce a state $\ket{r}^R  \in \H_B^R \subseteq \tH$ defined above, 
such that Bob processes and measures $\I_B \ket{r}^R= \ket{r}$, 
then she can get full information of Bob's measured outcomes without changing their statistics.
\end{theorem}
\textit{Proof.}
Eve captures the incoming state~$\ket{\varphi}$ and applies 
the same measurement as Bob on~$\I_B \imath\ket{\varphi}$, 
getting some $\ket{r}$ in~$\B$.  She then sends Bob $\ket{r}^R$. Bob applies
$\I_B $ to it which produces the state $\ket{r}$, 
i.e. Bob's complete measurement will yield~$\ket{r}$, the same value obtained by Eve.
\hfill$\square$

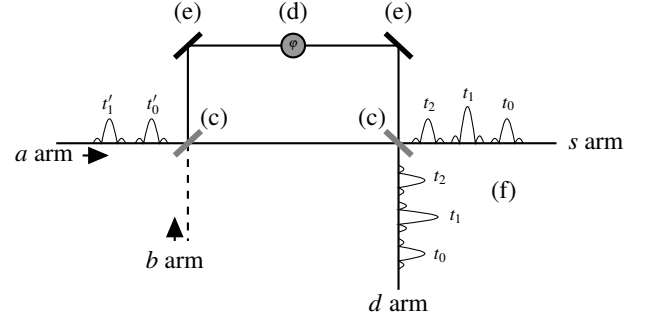
\begin{figure} 
 \centering 
\begin{tikzpicture}[xscale=0.7,yscale=0.65]

	\inter;

	\node at (phase) [label=above:{(d)}] {\tiny$\phi$};
	\node at ($(bs1)+(0.5,0.5)$) {(c)};
	\node at ($(bs2)+(-0.5,0.5)$) {(c)};
	
	\node at (9.75,0) {$s$ arm};
	\node at (6,-3.25) {$d$ arm};
	\node at (8,-1) {(f)};
	
	\node at (mir1) [label=above:{(e)}] {\phantom{\tiny 0}} ;
	\node at (mir2) [label=above:{(e)}] {\phantom{\tiny 0}};

	\draw[-triangle 45,   thick]  (0,-0.25) node[auto,left]{$a$ arm} -- (0.5,-0.25);
	\draw[-triangle 45,   thick]	 (1.75,-2)  node[auto,below] {$b$ arm}--(1.75,-1.5);

	\pulse{(0.2,0)}{0.5};
	\pulse{(1,0)}{0.5};
	\pulse{(6.25,0)}{0.5};\pulse{(7,0)}{0.75};\pulse{(7.75,0)}{0.5};	
	\rotatepulse{(6,-0.45)}{0.5};
	\rotatepulse{(6,-1.2)}{0.75};
	\rotatepulse{(6,-1.95)}{0.5};

	\node at (0.5,0.8) {\scriptsize $t'_1$};
	\node at (1.35,0.8) {\scriptsize $t'_0$};
	
	\node at (6.6,0.8) {\scriptsize $t_2$};
	\node at (8.1,0.8) {\scriptsize $t_0$};
	\node at (7.35,1) {\scriptsize $t_1$};
	
	\node at (6.8,-0.7) {\scriptsize $t_2$};
	\node at (7.1,-1.5) {\scriptsize $t_1$};
	\node at (6.8,-2.3) {\scriptsize $t_0$};

\end{tikzpicture}
 \caption
 {A Mach-Zehnder interferometer. 
  An input qubit enters from the $a$~arm while a vacuum ancilla is added by  %
  arm $b$.
 (c) beamsplitters; (d) phase shifter~$P_\phi$; 
 (e) mirrors;
 (f) six output modes.  
 }
 \label{fig:lab-xy} 
\end{figure}

\nsection{Application: an attack on Interferometric-based scheme}\label{sec:interferometer}
Our second interesting example for a fixed-apparatus setting which suffers from 
the fixed-apparatus attack is \emph{interferometric QKD}, realized via 
a Mach Zehnder interferometer (Fig.~\ref{fig:lab-xy}).
The most common interferometric realization is 
for BB84 with $x$ and~$y$ bases~\cite
{T94} (see also \cite{HMP00,GRTZ02,BBN03,DLH06}),
in which Bob changes the phase $\phi$ of the phase-shifter according to the basis he wishes to measure (see, e.g.,~\cite{GM12} for a detailed description of this setting).

However, 
motivated by the need to increase the key rate
and other reasons\footnote
	{Measuring the $z$ basis might be required, for instance, 
	in order to implement the 6-state QKD protocol~\cite{Brus98}, 
	in which  Alice sends a qubit using
	the $x$, $y$ and $z$ bases at random;
	or in order to perform  ``QKD with classical Bob''~\cite{BKM07,BGKM09,ZQLWL09} 
	in which one party is restricted to use only the (classical) $z$-basis, and either performs 	measurements
	in that basis or returns the qubits (unchanged) to the other party.},
\emph{fixed} implementations of BB84 with $x$ and $z$ bases 
were suggested and implemented as well~\cite{GRTZ02,WASST03,NHN03,NHN04,JS06}.
At each transmission, Alice sends a single photon in a superposition of two
time-modes $\ket{t'_0}\equiv \ket{0}$ and $\ket{t'_1}\equiv \ket{1}$.
Bob fixes {the phase shift}~$\phi$ to the $x$-basis and performs a complete measurement of the space. Later Alice reveals the
basis she used. When the $x$ basis is used Bob ignores clicks at time $t_0$ and~$t_2$ as those measurements are inconclusive. Similarly, when the $z$~basis is used, Bob ignores clicks at~$t_1$.

Now that the apparatus is fixed,  the implementation suffers from our attack (Theorem~\ref{thm:gen1}). 
As in the case of polarization-based BB84, 
one of the interferometer's input arms is assumed to be blocked, and if this is
not the case then the scheme is totally insecure:\footnote{It was shown that such a
scheme might not be fully
secure even when the second input arm is blocked~\cite{GM12}.}
each one of the states Bob may detect can
be ``reversed in time'' yielding a superposition
of single-photon states (of four modes at the interferometer inputs) that Eve can generate.

While the action of the interferometer can be mathematically defined on a %
space including pulses with more than a single photon, 
we only care about states with at most one photon.
In order to properly describe our setting, we consider  6 time-modes
corresponding to times $t'_{-2},\ldots,t'_3$ 
where each ``time mode'' contains either a single photon or zero photons and denote
$\ket{t'_n}_a$  the state with 1 photon at time $t'_n$ and $0$ at other times in arm $a$;
 $\ket{V}_a$  denotes the state with 0 photon at all 6 times.
The space $\tH_A$ is then taken to be the span of the states $\ket{V}_a, \ket{t'_{-1}}_a, \ldots, \ket{t'_{2}}_a$.
Similarly 
to what we saw in Section~\hyperref[sec:fattack]{II}, also here we let $\tH_\mathrm{anc}$ contain more states
than just the vacuum. It
is spanned by the states $\ket{V}_b, \ket{t'_{-1}}_b,\ldots, \ket{t'_{2}}_b$ of arm $b$.
Since we only need to care about pulses with at most a single photon we define the space $\tH$
to be the span of the states $\ket{V}_{ab} = \ket{V}_a\ket{V}_b$,
$\ket{t'_n}_a\ket{V}_b$ and $\ket{V}_a\ket{t'_n}_b$,  for $-1\leq n\leq 2$. %

Due to space constraints, we  use the following notations:
$\ket{a_n} \equiv \ket{t'_n}_a\ket{V}_b$, $\ket{b_n} \equiv \ket{V}_a\ket{t'_n}_b$;
similarly for the outputs at times $t_n$, we let $\ket{s_n} \equiv \ket{t_n}_s\ket{V}_d$ and $\ket{d_n} \equiv \ket{V}_s\ket{t_n}_d$.
The action of the interferometer on $\tH$ can then be described by the following equations:
\begin{equation}\label{eqn:U}
\begin{split}
\I_B{\ket{a_n}}   &=
 \phantom{\mathrm{i}}(\ket{s_n}-\ket{s_{n+1}}+\mathrm{i}\ket{d_n} +\mathrm{i}\ket{d_{n+1}})  / 2  \\ 
\I_B{\ket{b_n}}  &=
 \mathrm{i}(\ket{s_n} + \ket{s_{n+1}}+\mathrm{i}\ket{d_n} - \mathrm{i}\ket{d_{n+1}})/2
\ ,
 \end{split}
 \end{equation}
and $\I_B\ket{V}_{ab}= \ket{V}_{sd}$.
While $\tH$ 
and $\I_B(\tH)$ are of dimension~9,
the output states 
$\ket{V}_{sd}$, 
$\ket{s_{-1}}$, \!$\ket{s_0}$, \!$\ket{s_1}$, \!$\ket{s_2}$, \!$\ket{s_3}$, \!$\ket{d_{-1}}$, \!$\ket{d_0}$, \!$\ket{d_1}$, \!$\ket{d_2}$, \!$\ket{d_3}$
generate a space of dimension~11
(with some non zero occupancy numbers at time $t_3$ whereas all occupancy
numbers for $\ket{a_n}$ and $\ket{b_n}$ at times $t'_{-2}$ and $t'_3$ are $0$).
Recall that Bob measures the six basis states %
$\ket{s_0}$, \!$\ket{s_1}$, \!$\ket{s_2}$, \!$\ket{d_0}$, \!$\ket{d_1}$, \!$\ket{d_2}$
depicted in Fig.~\ref{fig:multiplex}.
Using Eq.~\eqref{eqn:U}
it is easy to check that for each state $\ket{r}$ of Table~\ref{table:Battack},
the equality $\I_B\ket{r}^R = \ket{r}$ holds. 
Bob's reversed space~$\H_B^R$ is the span of those states~$\ket{r}^R$.
If Eve is capable of generating and sending the states $\ket{r}^R$ she  fully
breaks the protocol (Theorem~\ref{thm:gen1}).  
For full details on how Table~\ref{table:Battack} is derived, see Section~\hyperref[app:int]{B} in the Supplemental Material.

\begin{table}[ht]
\begin{center}
\begin{tabular}{ccll}
$m$ &\hspace*{1em}& \textit{\textbf{$\phantom{\mathrm{i}}\ket{r}$}} & { $\qquad\qquad\quad\ket{r}^R$ } \\ \hline\hline
$(z,0)$ && $\phantom{\mathrm{i}}\ket{s_0}\quad$ & $(\ket{a_0} - \ket{a_{-1}} - \mathrm{i}\ket{b_0} - \mathrm{i}\ket{b_{-1}})/2$   \\
$(z,0)$ &&$\mathrm{i}\ket{d_0}$ &$(\ket{a_0} + \ket{a_{-1}} - \mathrm{i}\ket{b_0} + \mathrm{i}\ket{b_{-1}})/2$  \\
$(x,1)$ && $\phantom{\mathrm{i}}\ket{s_1}$ &$(\ket{a_1} - \ket{a_{0}} - \mathrm{i}\ket{b_1} - \mathrm{i}\ket{b_{0}})/2$  \\
$(x,0)$ &&$\mathrm{i}\ket{d_1}$ &$(\ket{a_1} + \ket{a_{0}} - \mathrm{i}\ket{b_1} + \mathrm{i}\ket{b_{0}})/2$  \\
$(z,1)$ && $\phantom{\mathrm{i}}\ket{s_2}$ & $(\ket{a_2} - \ket{a_{1}} - \mathrm{i}\ket{b_2} - \mathrm{i}\ket{b_{1}})/2$  \\
$(z,1)$ && $\mathrm{i}\ket{d_2}$ &$(\ket{a_2} + \ket{a_{1}} - \mathrm{i}\ket{b_2} + \mathrm{i}\ket{b_{1}})/2$  \\
\end{tabular}

\end{center}
\caption{Bob's measured states and their reversal.}
\label{table:Battack}
\end{table}

\nsection{IV. The Fixed-Apparatus Attack when Bob combines several outcomes}\label{sec:forget}
Table\,\ref{table:Battack} contains an additional column for $m$. 
When Bob measures either $\ket{s_0}$ or $\ket{d_0}$ (for implementing BB84), he draws the same conclusion:
``basis~$z$'' %
and ``bit $0$''. If after getting any of those two measurements  Eve  sends Bob 
the state $\ket{\psi_{z,0}} = \frac{1}{\sqrt{2}}[\ket{a_0} -\mathrm{i}\ket{b_0}]$, then $\I_B\ket{\psi_{z,0}} = \frac{1}{\sqrt{2}}[\ket{s_0}+\mathrm{i}\ket{d_0}]$
so Bob still measures
$\ket{s_0}$ or $\ket{d_0}$ with equal probability and gets the same result $(z,0)$. Similarly, if Eve  sends Bob the state 
$\ket{\psi_{z,1}} = \frac{1}{\sqrt{2}}[\ket{a_1}+\mathrm{i}\ket{b_1}]$
when she measures either $\ket{s_2}$ or $\ket{d_2}$, Bob measures $\ket{s_2}$ or $\ket{d_2}$ with equal probability and concludes the same result~$(z,1)$.

The attack of this section is simpler and more practical compared to the attack based on the $6$ states $\ket{r}^R$
of Table~\ref{table:Battack}
as now
all  pulses sent by Eve  are at times $t'_0$ and~$t'_1$, 
exactly the times used by Alice.
Note that if Alice and Bob modify their protocol and add detectors also on time slots $t'_{-1}$ and $t'_{-2}$ 
 (or add shutters on the input arm $a$) and verify statistics, this will help them to prevent Eve's attack of Section~\hyperref[sec:nontrivial]{III}, yet not that of Section~\hyperref[sec:forget]{IV}.

If $M$ is the set of outputs $m$, and $\mu:\B \to M$ is the function that
associates to each basis state the corresponding $m$ in $M$,
a general statement can be stated as follows:
\begin{theorem}\label{thm:forgetting}
If for all $m\in M$ 
Eve can produce a state $\ket{m}^R  \in \tH$ 
such that the state $\I_B \ket{m}^R$ is in 
the span of $\{\ket{r}\in \B \mid \mu(r)=m\}$,
then she can get full information of Bob's measured outcomes without changing their statistics.
\end{theorem}
This theorem assumes Bob checks only the outputs in $M$;  the statistics of individual basis measurements in the basis~$\B$,
in general, may not be preserved.

Note that a similar (and even simpler)
extension applies to the simple case of Theorem~\ref{thm:main}, in which Bob's setting is a unitary operator ($\U_B$) rather than an isometry.

\nsection{V. Conclusions}
Theorems~\ref{thm:main}--\ref{thm:forgetting} %
apply to several QKD experiments in which a fixed-apparatus is used by Bob, such as~\cite{BHKLLMNPS98,HNMP02,HNDP02,RTGK02,KZHWGTR02,ATMYRBBPG04,WASST03,NHN03,NHN04,JS06}.
We stress that we don't claim these implementations to be insecure. 
Rather, we point out weak-points of such realizations. Namely, they are insecure
only against a very strong eavesdropper.
A similar attack is potentially possible whenever the space Bob measures for the z-basis is orthogonal to the space he measures for the x-basis (which is always the case in fixed-apparatus schemes).

The attacks we present require Eve to have one-time access to Bob's device, 
e.g., to drill a small hole in Bob's device, right where
the other input of the beamsplitter (assumed to be blocked) is located; 
or to wire the other arm into the channel, using some time-multiplexing.
If Eve is some technician or service person that has one time access to Bob's lab, 
or if Alice and Bob purchase their devices from Eve\footnote
{Of course, if Eve \emph{builds} the device she has a lot of power and can potentially perform much stronger attacks.},
the device may be compromised.

\textbf{Acknowledgements.}
RG is grateful to Technion, Israel for hosting him while part of this work was done.
The work of TM was supported in part by the Wolfson Foundation, by the Israeli MOD Research and Technology Unit, by FQRNT through INTRIQ, and by NSERC.
The work of MB was supported in part by FQRNT through INTRIQ, and by NSERC.
 


%


\section*{Supplemental Material}

\section{A. Attacks on Fixed-Apparatus Polarization-Based BB84}\label{app:BS}
In this section we demonstrate the power of the reversed attack on a fixed apparatus by devising an
attack on a very simple (yet common) setup of the BB84 scheme, based on polarized photons. 
We begin by describing the setup and the protocol over the setup and then move on to describing Eve's attack.

\subsection{1. Polarized photons and Fock notations}
In polarization-based BB84,
each qubit is implemented via a single photon which either has  a horizontal/vertical polarization 
(the {$+$} basis with basis states $\ket{\zp}$ and~$\ket{\op}$),
or  a diagonal polarization, i.e., $\pm45^\circ$ 
(the  {{$\times$}}  basis, with basis states $\ket{\zd}$ and~$\ket{\od}$).

With the Fock state notations (which we denote using a ket with an `{\footnotesize{F}}' superscript), 
a polarized photon is described using two modes corresponding to 
orthogonal polarization states, where the first mode corresponds to horizontal polarization
and the second mode to vertical one.
Hence, $\ket{\zp} \equiv \fet{10}$ (one horizontally polarized photon, 
zero vertically polarized photon), and $\ket{\op}\equiv \fet{01}$.
An arbitrary polarization state $\ket{\phi}$ 
of a single photon is 
written using the computation basis as  
$\ket{\phi} = \alpha \ket{\zp} + \beta \ket{\op}$ or using Fock notations as
$\fet{\phi}=\alpha \fet{10} + \beta \fet{01}$.
In addition to the case of a pulse with a single photon described above, 
we use $\ket{V}\equiv\fet{00}$ to denote a pulse with zero photons (a vacuum state).

While it is important
to keep the Fock notations in mind, they become a bit cumbersome. We
can actually do well without them in this section
because the physical situation we deal with
is relatively simple. 
In the following we use both notations interchangeably 
to describe Bob's apparatus, 
however we will use the standard notations for describing 
the relevant spaces and our fixed-apparatus attack on this setting.

\subsection{2. Bob's setup}

Bob's setup contains beam splitters of two types, polarization rotators, and
detectors. 
See Fig.~\ref{fig:multiplex}. We now derive the transformation $\calu_B$ induced by his equipment.

The matrix describing the unitary 
transformation applied by a $50\%/50\%$ beam splitter (BS) is 
\begin{equation}\label{eqn:BS}
\mathrm{BS} = \frac{1}{\sqrt{2}}\begin{bmatrix}1 & \mathrm{i} \\ \mathrm{i} & 1\end{bmatrix}
\end{equation}
meaning that a transmitted particle gets no phase and a reflected particle gets
the phase $\mathrm{i}$.
This implies 
the following transformation for a single particle:
\begin{align*}
\fet{1}_a\fet{0}_b &\mapsto
\frac{1}{\sqrt{2}}\big[\phantom{\mathrm{i}}\fet{1}_c\fet{0}_d +
\mathrm{i}\fet{0}_c\fet{1}_d\big]\\
\fet{0}_a\fet{1}_b &\mapsto
\frac{1}{\sqrt{2}}\big[\mathrm{i}\fet{1}_c\fet{0}_d +
\phantom{\mathrm{i}}\fet{0}_c\fet{1}_d\big]
\end{align*}
where $a$ and~$b$ are the modes entering the beamsplitter and $c$ and~$d$ are the output modes (see, e.g., the PIBS in Fig.~\ref{fig:multiplex}).

Next, we extend the discussion to also deal with the polarization of the single particle.
A BS that does not change polarization (a polarization-independent BS --- PIBS)
applies the transformation of Eq.~\eqref{eqn:BS} to either horizontally polarized photon or
vertically polarized photon, hence to any polarization state~$\ket{\phi}$.
With Fock-state notations the PIBS acts as follows
 on a single photon having a known polarization state:
\begin{align*}
\text{PIBS} \fet{\phi}_a\fet{00}_b &= \frac{1}{\sqrt{2}}\Big[\phantom{\mathrm{i}}\fet{\phi}_c\fet{00}_d + \mathrm{i}\fet{00}_c\fet{\phi}_d\Big]\\
\text{PIBS}\fet{00}_a\fet{\phi}_b &= \frac{1}{\sqrt{2}}\Big[\mathrm{i}\fet{\phi}_c\fet{00}_d + \phantom{\mathrm{i}}\fet{00}_c\fet{\phi}_d\Big],
\intertext{and with standard qudit notation this is written as}
\text{PIBS}\ket{\phi}_a\ket{V}_b &= \frac{1}{\sqrt{2}}\Big[\phantom{\mathrm{i}}\ket{\phi}_c\ket{V}_d + \mathrm{i}\ket{V}_c\ket{\phi}_d\Big]\\
\text{PIBS}\ket{V}_a\ket{\phi}_b &= \frac{1}{\sqrt{2}}\Big[\mathrm{i}\ket{\phi}_c\ket{V}_d + \phantom{\mathrm{i}}\ket{V}_c\ket{\phi}_d\Big].
\end{align*}

Bob adds a polarization rotator $\mathrm{PR}$ on the $d$~arm, 
whose operation on the vacuum is the identity $\mathrm{PR}\ket{V} = \ket{V}$, and
on diagonally polarized states is 
\[
\mathrm{PR}\ket{{\zd}} = \ket{\zp} \quad ; \quad
\mathrm{PR}\ket{{\od}} = e^{\mathrm{i}\theta}\ket{\op}  . 
\]
The phase~$\theta$  is arbitrary and may be unknown to all parties
since it results in a global phase when Bob measures.

Therefore, Bob's transformation~$\U_B$ 
from the input arms to just before the measuring apparatus satisfies
\begin{equation}\label{eqn:POL}
\begin{split}
\U_B\ket{\phi}_a\ket{V}_b &= \frac{1}{\sqrt{2}}\Big[\phantom{\mathrm{i}}\ket{\phi}_c\ket{V}_d + \mathrm{i}\ket{V}_c\mathrm{PR}\ket{\phi}_d\Big]\\
\U_B\ket{V}_a\ket{\phi}_b &= \frac{1}{\sqrt{2}}\Big[\mathrm{i}\ket{\phi}_c\ket{V}_d + \phantom{\mathrm{i}}\ket{V}_c\mathrm{PR}\ket{\phi}_d\Big].
\end{split}
\end{equation}

Finally, the state reaches Bob's 
 measuring apparatus  which
contains a polarization beam splitter (PBS) and a
detector on each arm (for each of the two bases). 
A PBS is such that a $\ket{\zp}$ (horizontally polarized) photon goes through
towards one detector, 
and a $\ket{\op}$ (vertically polarized) photon is reflected towards the other
detector. As before, the transmitted photon gets no phase while the reflected photon gets an additional  phase~$\textrm{i}$. 

For simplicity we
assume that Bob's detectors can distinguish zero photons, from one photon,
and from
more than one, namely these detectors are perfect counters.
As a result, the protocol works fine when one and only one
of Bob's detectors finds exactly one photon.

\subsection{3. The embedding of $\mathcal{H}_A$ and the relevant spaces}
The embedding of 
$\mathcal{H}_A$ into the (a--b) system is given by
$
\mathcalus{I}\ket{\phi} = \fet{\phi}_a\fet{00}_b
$, or without fock-state notations, 
$\mathcalus{I}\ket{\phi} = \ket{\phi}_a\ket{V}_b$.
Alice's states are actually embedded into larger spaces, but our aim
here is to focus on the
minimal space required for proving the insecurity of the protocol. 
Let us specify a relevant embedding space, 
the two polarization modes of arm~$a$ limited to up to a single photon, 
which span a space {of dimension $3$} which we denote~$\mathcal{\widetilde H}_A$, 
and the additional two ancilla modes in Bob's
other input~$b$ limited to up to a single photon, denoted~$\tH_{\mathrm{anc}}$.
The entire relevant embedding 
space is thus 
the $9$ dimensional space~$\tH_A \otimes \tH_{\mathrm{anc}}$.

If we try to choose $\tH_A$ or $\tH_\text{anc}$ that contain less states,
then $\H_B$ will be outside the space resulting from the 
application of Bob's setup onto Alice's states.
This is more easily seen by the ``reversed-space'' analysis:  
if we begin with a photon in one of Bob's detector and 
`reverse' the direction the photon travels 
(i.e., assume it goes from the detector and to the channel), 
it is possible that the photon will go to arm~$b$ leaving a vacuum state 
in arm~$a$. However, $\ket{V}_a\notin \H_A$ and we must choose 
$\tH_A$ to include also 
a vacuum state.
By checking the reversed transformation on the four outcome states, we see
that $\tH_\text{anc}$ must also contain three states, the vacuum and the two
single-photon polarization states. 

We only need to care about 
pulses with at most a single photon,
thus  
we can let $\tH$ be the span of
$\{\ket{\phi}_a\ket{V}_b , 
\ket{V}_a\ket{\phi}_b , \ket{V}_a\ket{V}_b\}$ for $\ket{\phi}\in \{\ket{\zp}_a, \ket{\op}_a\}$.
Thus, $\tH$ is 
a five dimensional subspace of the 9-dimensional space 
$\tH_A\otimes\tH_\mathrm{anc}$.
Note that Bob's transformation $\calu_B$ is well defined on~$\tH$ using Eq.~\eqref{eqn:POL}, along with
$\calu_B\ket{V}_a\ket{V}_b = \ket{V}_c\ket{V}_d$.

\subsection{4. Bob's reversed space and Eve's attack}

Recall that Bob's measured space~$\H_B$ is defined to be the span of the 4 linear-independent states
that cause his four detectors to  click with certainty. 
Those are, 
\begin{table}[H]
\begin{center}
\renewcommand\arraystretch{1.2}
\begin{tabular}{ccc}
\hline
 \parbox[t]{2,5cm}{\centering \textbf{\textit{Detector clicking}}} & \parbox[t]{2.5cm}{\centering \textbf{\textit{State when arriving at PBS}}} & 
\parbox[t]{2cm}{\centering \textbf{\textit{State just after PIBS}}}  \\ \hline
\papl{\mathtt{+\,0}} & $\phantom{e^{\mathrm{i}\theta}}\ket{\zp}_{c}\ket{V}_{d}$ & $\ket{\zp}_{c}\ket{V}_{d}$ \\
\papl{\mathtt{+\,1}} & $\phantom{e^{\mathrm{i}\theta}}\ket{\op}_{c}\ket{V}_{d}$ & $\ket{\op}_{c}\ket{V}_{d}$ \\
\papl{\mathtt{\times\, 0}} & $\phantom{e^{\mathrm{i}\theta}}\ket{V}_{c}\ket{\zp}_{d}$ & $\ket{V}_{c}\ket{\zd}_{d}$ \\
\papl{\mathtt{\times\, 1}} & $e^{\mathrm{i}\theta}\ket{V}_{c}\ket{\op}_{d}$ & $\ket{V}_{c}\ket{\od}_{d}$ \\
\end{tabular}
\end{center}
\caption{Bob's measured space $\H_B$}
\end{table}
\noindent

In order to get Bob's reversed space, 
we need to apply the complex conjugate of the $\mathrm{PIBS}$ matrix to 
the four orthogonal states in the right-hand column (state just after PIBS) 
to get states in the input of the (a--b) system that
are relevant for Eve's attack:
\begin{table}[H]
\begin{center}
\renewcommand\arraystretch{1.2}
\begin{tabular}{cc} \hline \parbox[t]{3cm}{\centering\textit{\textbf{ Detector
clicking}}} & \parbox[t]{5cm}{\centering\textit{\textbf{Input state forcing the click}}}\\ \hline
\papl{\mathtt{+\,0}} & $\frac{1}{\sqrt{2}}\big[\ \phantom{-\mathrm{i}}\ket{\zp}_{a}\ket{V}_{b} -\mathrm{i}\ket{V}_{a}\ket{\zp}_{b}\big]$\\
\papl{\mathtt{+\,1}} & $\frac{1}{\sqrt{2}}\big[\ \phantom{-\mathrm{i}}\ket{\op}_{a}\ket{V}_{b} -\mathrm{i}\ket{V}_{a}\ket{\op}_{b}\big]$\\
\papl{\mathtt{\times\, 0}} & $\frac{1}{\sqrt{2}}\big[-\mathrm{i}\ket{\zd}_{ a}\ket{V}_{b} +\phantom{\mathrm{i}}\ket{V}_{a}\ket{\zd}_{ b}\big]$\\
\papl{\mathtt{\times \,1}} & $\frac{1}{\sqrt{2}}\big[-\mathrm{i}\ket{\od}_{ a}\ket{V}_{b} +\phantom{\mathrm{i}}\ket{V}_{a}\ket{\od}_{ b}\big]$\\
\end{tabular}
\end{center}
\caption{Bob's reversed measured space~$\H^R_B$}
\end{table}
The above states are those that span the space~$\H_B^R$ available for the eavesdropper attack.
It is easy to verify that indeed $\H_B^R \subseteq \tH$, since each of the above states is in the span of $\{
\ket{\phi}_a\ket{V}_b , 
\ket{V}_a\ket{\phi}_b , \ket{V}_a\ket{V}_b
\}$, for $\ket{\phi}$ a polarization qubit. This validates our choice for~$\tH$, and also suggests that we could have omitted~$\ket{V}_a\ket{V}_b$ and obtain a smaller~$\tH$.


\section{B. Attacks on Fixed-Apparatus Interferometric BB84}\label{app:int}
In this section we devise two fixed-apparatus attacks on a BB84 implementation
based on a Mach-Zehnder interferometer. 
We begin by describing the setting and deriving the transformation induced by Bob's apparatus.
Next we show a fixed apparatus attack that uses Bob's entire reversed space (following Theorem~\ref{thm:gen1}).
Finally, we devise an even stronger fixed-apparatus attack (following Theorem~\ref{thm:forgetting}), which is restricted to a subspace of Bob's reversed space which corresponds to ``valid'' signals received by Alice.\looseness=-1

\subsection{1. The interferometer}
\begin{figure}[ht] 
 \centering 
\begin{tikzpicture}[scale=0.9,xscale=0.8]

	\inter;
	
	\node at (phase) [label=above:{(d)}] {\tiny$\pi$};
	\node at ($(bs1)+(0.5,0.5)$) {(c)};
	\node at ($(bs2)+(-0.5,0.5)$) {(c)};
	
	\node at (9.75,0) {$s$ arm};
	\node at (6,-3.25) {$d$ arm};
	\node at (8.5,-1) {(f)};
	
	\node at (mir1) [label=above:{(e)}] {\phantom{\tiny 0}} ;
	\node at (mir2) [label=above:{(e)}] {\phantom{\tiny 0}};
	
	\draw[-triangle 45,   thick]  (-0.5,-0.25) node[auto,left]{(a)} -- (0,-0.25);
	\draw[-triangle 45,   thick]	 (1.75,-2)  node[auto,below] {(b)}--(1.75,-1.5);
	
	\pulse{(0.2,0)}{0.5};
	\pulse{(1,0)}{0.5};
	\rotatevacuum{(2.1,-0.5)}{0.5};
	\rotatevacuum{(2.1,-1.3)}{0.5};

	\pulse{(6.25,0)}{0.5};\pulse{(7,0)}{0.75};\pulse{(7.75,0)}{0.5};	
	\rotatepulse{(6,-0.45)}{0.5};
	\rotatepulse{(6,-1.2)}{0.75};
	\rotatepulse{(6,-1.95)}{0.5};

	\node at (0.5,0.8) {\small $a_1$};
	\node at (1.35,0.8) {\small $a_0$};
	
	\node at (3,-0.8) [black!50] {\small $b_0$};
	\node at (3,-1.6) [black!50] {\small $b_1$};

	\node at (6.6,0.8) {\small $s_2$};
	\node at (8.1,0.8) {\small $s_0$};
	\node at (7.35,1) {\small $s_1$};
	
	\node at (7,-0.7) {\small $d_2$};
	\node at (7.15,-1.5) {\small $d_1$};
	\node at (7,-2.3) {\small $d_0$};

	\draw[|<->|] (0.5,-0.2) -- node[below]{$\Delta T$} (1.3,-0.2);
	\draw[|<->|] (5.8,-1.525) -- node[left]{$\Delta T$} (5.8,-2.25);
	\node at ($(phase)-(0,0.5)$) {\small $\Delta T$ detour};

\end{tikzpicture}
 \caption
 {A Mach-Zehnder interferometer. 
 (a) An input qubit. The time-difference between the two incoming modes is 
identical to the difference between the two arms; 
 (b) a vacuum state entering the second arm;
 (c) beamsplitters; 
 (d) $\Delta T$ time detour plus a phase shift (to compensate phase shift done by mirrors);
 (e) mirrors;
 (f) six output modes. 
 }
\label{fig:lab-xy:app} 
\end{figure}
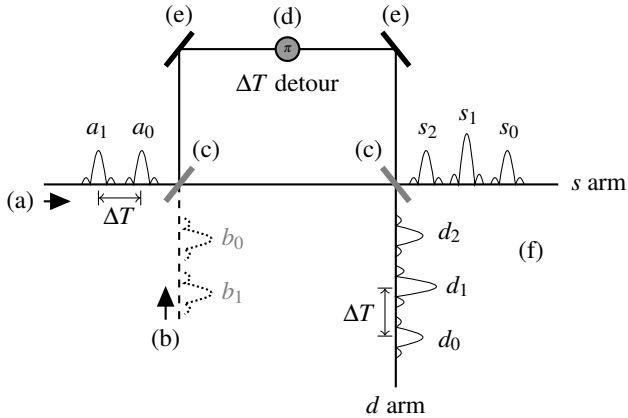

Consider a BB84 implementation which uses 
two time-separated modes (Fig.~\ref{fig:lab-xy:app}).
For every transmission, the first mode arrives to Bob's lab at time~$t'_0$, and the second
mode at $t'_1=t'_0+\Delta T$ on arm~$a$. 
As in the previous section, we can describe the number of particles in each
mode using Fock notations, $\fet{10}_a$ for a photon at time $t'_0$ and $\fet{01}_a$ for a photon at time $t'_1$.

Assume a photon enters the interferometer at time $t'_0$
(i.e., the state $\fet{10}_a$). 
After some fixed time~$T$ (i.e., at time $t_0 \equiv t'_0+T$), 
if it goes straight~($s$) through the two beam splitters,
it creates a pulse at the $s$ output arm, and if reflected by the second
BS it goes down~($d$) and produces a pulse at the $d$ output arm.
We also assume that the upper detour causes a
delay equal to~$\Delta T$,
thus if the photon is reflected by the first BS, it will produce a 
pulse at time $t'_0+T+\Delta T$ at either the $s$ or $d$ output arms.
We denote the output times modes as $t_0=t'_0+T$ and $t_1=t'_1+T$, etc.
Note that due to the precise timing between consecutive input time modes and the length of the detour in the interferometer, $t'_0+T+\Delta T = t_1$, which causes each output mode to be the interference of two consecutive input time modes.

It follows that a photon in states $\fet{10}_a, \fet{01}_a$ can end out in one of 6 possible modes after the interferometer, these correspond to times $t_0,t_1,t_2$ in the $s$ arm and in the $d$ arm denoted as $\fet{p_0p_1p_2}_s\fet{q_0q_1q_2}_d$ where $p_n, q_n$ are integers denoting the occupancy of time $t_n$ in the respective arm.
Finally, it is known that every reflection on a beam splitter (BS) produces a phase
factor equal to $\textrm{i}$. For instance, if $\fet{10}_a$ goes through the first BS and then is reflected by
the second BS, it ends in state~$\mathrm{i}\fet{000}_s\fet{100}_d$; it can also be reflected by the first BS (phase factor~$\mathrm{i}$), be delayed $\Delta T$ 
by the upper detour, go throughout the second BS (phase factor~$1$) and then end in state~$\mathrm{i}\fet{000}_s\fet{010}_d$. Note that an additional phase of $\mathrm{i}$ is added each time the photon is reflected from a mirror. To compensate for these extra phases we add a (fixed) phase-shift of $\pi$ and neglect the effect of the mirrors hereinafter.

In a BB84 implementation that uses the $x$ and $z$ bases, Alice and Bob encode the states as follows:
\begin{align*}
\ket{0_z} \equiv  \fet{10}_a &&&
\ket{0_x}  \equiv  \left ( \fet{10}_a + \fet{01}_a \right)/\sqrt{2} \phantom{\text{ .}}\\ 
\ket{1_z}  \equiv  \fet{01}_a &&&
\ket{1_x}  \equiv  \left ( \fet{10}_a - \fet{01}_a \right )/\sqrt{2}\text{ .}
\end{align*}
We complete the description of the protocol (namely, how the above states evolve through the interferometer and measured by Bob) after 
discussing the spaces $\tH_A$, $\tH_\anc$, and~$\tH$, and the inclusion~$\I$ of $\H_A$ into $\tH_A\otimes\tH_\anc$.
Then, we define~$\I_B$, the interferometer transformation on~$\tH$ (which also includes $\H_A$), and explain Bob's measurement process.

\subsection{2. The embedding of $\H_A$ and the relevant spaces}
As explained above, the states sent by Alice are in the space $\H_A$ spanned by 
$\{\fet{10}_a$, $\fet{01}_a\}$
where the Fock occupancy numbers
are for times $t'_0$ and $t'_1$ and the arm is the $a$~arm. However,
a simple analysis of Fig.~\ref{fig:lab-xy:app} shows that a photon entering at time~$t'_{-1}$ might also end up at time mode~$t_0$ at either the $s$ or~$d$ arms; Similarly, 
a photon at input time~$t'_2$ may also end up at output time mode $t_2$ in either arm.
Moreover, the way the interferometer evolves pulses at time mode~$t'_{n}$ depends on the pulses of times $t'_{n+1}$ and~$t'_{n-1}$. Thus, in order to understand how a (non--empty) pulse at time $t'_{-1}$ evolves through the interferometer, we must additionally consider time mode~$t'_{-2}$. Similarly, since our analysis uses non-empty pulses  time mode $t'_2$ we will need to add the time mode~$t'_3$. We remark that time modes $t'_{-2}$ and~$t'_{3}$ will \emph{always be empty}, i.e., contain a vacuum state, however they are needed for the validity of the unitary map.

Let us now define the embedding $\I= \I_\anc\I_A$ of Alice's space into $\tH_A\otimes \tH_\anc$.
Alice's inputs are  embedded into~$\tH_A$, 
the span of 
\begin{align*}
&\fet{010000}_a\equiv\ket{t'_{-1}}_a,  && \fet{001000}_a\equiv\ket{t'_{0}}_a, \\
&\fet{000100}_a\equiv\ket{t'_{1}}_a, && \fet{000010}_a\equiv\ket{t'_{2}}_a, \text{ and} \\
&\fet{000000}_a\equiv\ket{V}_a,
\end{align*}
where
the Fock occupancy numbers in this case correspond to times $t'_{-2}$ to $t'_3$ respectively in the $a$~arm: e.g.,  $\fet{010000}_a$ is one photon at time~$t'_{-1}$ in the $a$~arm
and $\fet{000010}_a$ is a photon at time~$t'_2$.
Alice inclusion is then given by
\[
\I_A\fet{10}_a = \ket{t'_0}_a \quad\text{ and }\quad \I_A\fet{01}_a = \ket{t'_1}_a.
\]

The ancillary space $\tH_\mathrm{anc}$, given by the $b$~arm, is similarly defined
as the span of 
\begin{align*}
&\fet{010000}_b\equiv\ket{t'_{-1}}_b,  && \fet{001000}_b\equiv\ket{t'_{0}}_b, \\
 &\fet{000100}_b\equiv\ket{t'_{1}}_b, && \fet{000010}_b\equiv\ket{t'_{2}}_b, \text{ and } \\
&\fet{000000}_b\equiv\ket{V}_b,
\end{align*}
with the same convention on times.
The inclusion $\I_\anc$ merely describes the addition of an ancillary vacuum state in the $b$ arm, thus
for  $n\in \{0,1\}$ we have $\I_\anc\ket{t'_n}_a = \ket{t'_n}_a\ket{V}_b$ (this can be extended to $-1\le n \le 2$).

In the following, the states $\ket{a_n}, \ket{b_n} \in \tH_A\otimes\tH_{\anc}$ 
are defined as  $\ket{a_n} = \ket{t'_n}_{a}\ket{V}_{b}$, and 
$\ket{b_n}=\ket{V}_{a}\ket{t'_n}_{b}$. 
We shall take $\tH$ to be the span of all the states $\ket{a_n}$ and $\ket{b_n}$ with
$-1 \le n \le 2$, as these are enough to span Bob's reversed space (Table~\ref{table:Battack}).

The interferometer induces a unitary embedding~$\I_B$  defined on~$\tH$; 
it takes $12$ input modes 
(where times $t_{-2}$ and~$t_3$ being systematically void)
and thus produces $12$ output modes.
The output is
embedded in a space   spanned by all the states of at most a single photon in one of the following $12$ modes: times $t_{-2}$ to $t_3$ in the $s$~arm and time $t_{-1}$ to $t_3$ in the $d$~arm. 
Similar to the naming convention above, we use the states $\ket{s_n}$ and~$\ket{d_n}$ with $-2\le n \le 3$ to denote a single photon at time~$t_n$ in the $s$ and $d$ arms respectively, and no photons in all other modes. For instance, the state~$\ket{s_{3}}$ represents one photon at time~$t_3$ in the $s$~arm, and vacuum in all other modes; $\ket{d_0}$ has a photon at time~$t_0$ in the $d$~arm, and vacuum everywhere else.

Recall that Alice and Bob use  the $x$ and $z$ bases encoded as follows:
\begin{align*}
\I\ket{0_z} =  \ket{a_0} &&&
\I\ket{0_x} =  \left ( \ket{a_0} + \ket{a_1} \right)/\sqrt{2} \phantom{\text{ .}}\\ 
\I\ket{1_z} = \ket{a_1} &&&
\I\ket{1_x} =  \left ( \ket{a_0} - \ket{a_1} \right )/\sqrt{2}\text{ .}
\end{align*}

Taking the four possible paths
for a pulse in the interferometer,
and taking into account the delay and the beamsplitter's phase factors,  gives the following
resulting states 
\begin{equation}\label{eqnB+-On01}
\begin{split}
\I_B\ket{a_0}   &=
 \left(\ket{s_0}-\ket{s_1}\phantom{{}-\ket{s_2}}
  +\mathrm{i}\ket{d_0}	+\mathrm{i}\ket{d_1} \phantom{{}+\mathrm{i}\ket{d_2}}\right)  / 2 \\  
\I_B \ket{a_1}   &=
 \left(\phantom{{}-\ket{s_0}} \ket{s_1}-\ket{s_2}
  \phantom{{}+\mathrm{i}\ket{d_0}}+\mathrm{i}\ket{d_1}+\mathrm{i}\ket{d_2}\right) / 2,
\end{split}
\end{equation}
which correspond to the output of the interferometer when Alice sends $\ket{0_z}$ and~$\ket{1_z}$, respectively.
The output of the interferometer when Alice sends $\ket{0_x}$ and $\ket{1_x}$ is the following linear combinations
of the right-hand side of~\eqref{eqnB+-On01}:
\begin{align}
\label{eqnB+-On+-}
\I_B\I\ket{0_x}  &= 
 \left(\ket{s_0}  \phantom{{}-2\ket{s_1}} -\ket{s_2} 
 +\mathrm{i}\ket{d_0} +2\mathrm{i} \ket{d_1} +\mathrm{i}\ket{d_2}\right)\!  /\! \sqrt{8} \nonumber \\
\I_B\I\ket{1_x}  &=  
 \left(\ket{s_0} -2\ket{s_1} +\ket{s_2} 
+\mathrm{i}\ket{d_0} \phantom{{}+2i\ket{d_1}} -\mathrm{i}\ket{d_2}\right)\!  /\! \sqrt{8}  .
\end{align}

In order to perform a measurement, Bob 
looks at his detectors at times $t_0, t_1$ and $t_2$, and writes down the mode that is occupied, that is,
he measures the basis $\B = \big\{\ket{s_0}$, $\ket{s_1}$, $\ket{s_2}$, $\ket{d_0}$, $\ket{d_1}$, $\ket{d_2}\big\}$.
Bob then interprets his measurement in a manner depicted in the following Table~\ref{table:mu}.
\begin{table}[H]
\begin{center}
\renewcommand\arraystretch{1.2}
\begin{tabular}{ll}\hline\textbf{\textit{State measured}\phantom{XXXX}} & \textbf{\textit{Meaning}}   \\ \hline
$\ket{s_0}$ or $\ket{d_0}$ & $(z,0)$     \\
$\ket{s_2}$ or $\ket{d_2}$ & $(z,1)$     \\
$\ket{d_1}$                & $(x,0)$     \\
$\ket{s_1}$                & $(x,1)$  
\end{tabular}
\caption{Bob's interpretation of measured state}\label{table:mu}
\end{center}
\end{table}

Bob's measurement space is thus $\H_B = \mathrm{Span}(\B)$. 
From equations \eqref{eqnB+-On01} and \eqref{eqnB+-On+-}, 
The probabilities of Bob getting each output on each state received from 
Alice are given in Table~\ref{table:stats}.
\begin{table}[H]
\begin{center}
\renewcommand\arraystretch{1.2}
\begin{tabular}{c|cccc} & $(z,0)$ & $(z,1)$ & $(x,0)$ & $(x,1)$ \\ \hline
$\ket{0_z}$ & $1/2$ & $0$ & $1/4$ & $1/4$ \\
$\ket{1_z}$ & $0$ & $1/2$ & $1/4$ & $1/4$ \\
$\ket{0_x}$ & $1/4$ & $1/4$ & $1/2$ & $0$\\
$\ket{1_x}$ & $1/4$ & $1/4$ & $0$ & $1/2$ \\
\end{tabular}
\end{center}
\caption{The probability that Bob measures each outcome in the interferometric BB84 for each state
sent by Alice}
\label{table:stats}
\end{table}

Consequently, if Alice encoded her bit in  basis $b\in \{z, x\}$ and Bob measures $(b,q)$ then they share bit $q$; if Bob does not
get the basis used by Alice, the bit he gets is equally likely $0$ or $1$.
\subsection{3. A fixed apparatus attack}
If Eve controls the $b$~arm, the above realization is totally insecure. 
If that input is left open, the apparatus is completely
symmetric; however, since a reflection causes a phase factor of $\mathrm{i}$, the reverse direction gives a factor of~$-\mathrm{i}$. 
Taking the detour back also decreases time instead of increasing it. 
Thus, if Eve wants to force Bob to measure, say, $\ket{s_1}$,  
she only needs to set as input of Bob's apparatus the state
\[
 (\ket{a_1}-\ket{a_0}-\mathrm{i}\ket{b_1} 	-\mathrm{i}\ket{b_0})  / 2 \text{.}
\]
That state
 can be calculated in the same way as \eqref{eqnB+-On01} but going backwards in Fig.~\ref{fig:lab-xy:app}, starting 
 from $\ket{s_1}$.

More generally, 
rewriting equations~\eqref{eqnB+-On01}  for an arbitrary time $-1 \le n \le 2$ gives
the unitary embedding $\I_B$ on the entire $\tH$:
\begin{equation}\label{eqn:U}
\begin{split}
\I_B{\ket{a_n}}   =
 (\phantom{i}\ket{s_n}-\phantom{i}\ket{s_{n+1}}+i\ket{d_n}
	+\mathrm{i}\ket{d_{n+1}})  / 2  \\ 
\I_B{\ket{b_n}}  =
 (i\ket{s_n} + \mathrm{i}\ket{s_{n+1}}-\phantom{i}\ket{d_n} + \phantom{i}\ket{d_{n+1}})/2
 \end{split}
 \end{equation}
 (in addition to the case where vacuum goes to vacuum).
The map $\I_B^\dagger$ is defined on $\B$ (and thus $\H_B$) by the formulas
\begin{align}\label{eqn:Udag}
\begin{split}
\I_B^\dagger\ket{s_n}   &= 
 (\phantom{-i}\ket{a_n}-\phantom{i}\ket{a_{n-1}}-i\ket{b_n}
	-i\ket{b_{n-1}})  / 2  \\
\I_B^\dagger\ket{d_n}  &= 
 (-i\ket{a_n} - i\ket{a_{n-1}}-\phantom{i}\ket{b_n} + \phantom{i}\ket{b_{n-1}})/2.
 \end{split}
 \end{align}

One can easily verify
for any $\ket{r}\in \B$ that the state $\ket{r}^R= \J_B\ket{r}$ given by Eq.~\eqref{eqn:Udag}
satisfies $\I_B\ket{r}^R = \ket{r}$ (thus for any~$\ket{r}\in \H_B$, it holds that
$\I^{\phantom{\dagger}}_B \J_B\ket{r} = \ket{r}$).
It follows that an eavesdropper can  force the  measurements $\ket{r}$
by simply inputting the  states $\ket{r}^R$ 
given in Table~\ref{table:Battack}, and thus fully breaking the protocol. 
These states also span Bob's reversed space~$\H_B^R$.


\subsection{4. Restricting inputs to $\tH_{\{0,1\}}$ i.e. times $0$ and $1$}

Bob has a very simple counter-measure for the above attack, which we now explain.
Assume Eve measures Alice's state using the basis $\B$ and gets~$\ket{d_0}$. 
She now needs to generate the state 
$(-\mathrm{i}\ket{a_0} - \mathrm{i}\ket{a_{-1}} - \ket{b_0} + \ket{b_{-1}})/2$ to force a measurement of~$\ket{d_0}$.
However, this  superposition uses pulses at  time $t'_{-1}$, \emph{which are not used in the original protocol}. 
Therefore, if Bob adds a shutter that blocks any inputs to his device except for times $t'_0$ and $t'_1$, 
he overcomes this attack.

  We  now put a more stringent condition on the inputs: they need to be in 
  $\tH_{\{0,1\}} = \mathrm{Span}(\ket{a_0}, \ket{a_1}, \ket{b_0}, \ket{b_1})$, i.e.\@
  only occupy times $t'_0$ and~$t'_1$. To allow an attack, we use the assumption
  that Bob does not check the individual outputs in~$\B$ but rather only takes into account the relevant outputs 
  for the protocol, as given in Table~\ref{table:mu}. The reason being that if Alice sends $\ket{0_z}$ or $\ket{1_z}$
  and Bob measures $(z,0)$ or $(z,1)$ then they share a bit. Similarly if Alice sends $\ket{0_x}$ or~$\ket{1_x}$ and Bob gets output $(x,0)$ or~$(x,1)$ then they share a bit. It is also easy to check that the probability that Bob gets the same basis as Alice is $1/2$ and when the bases are different, the bit obtained by Bob is random so that this protocol closely corresponds to the BB84 protocol.

Clearly $\tH_{\{0,1\}} \subseteq \tH$ and thus $\I_B$ is defined on $\tH_{\{0,1\}}$. 
Moreover, measurements $(x,0)$ and $(x,1)$ correspond to basis
states $\ket{d_1}$ and $\ket{s_1}$ and can only be produced inputting the states $\J_B\ket{d_1}$ and $\J_B\ket{s_1}$ (with a global phase) found
previously. Those are already elements of~$\tH_{\{0,1\}}$.

Measurement $(z,0)$ my be obtained by inputing any superposition of $\J_B\ket{s_0}$ and $\J_B\ket{d_0}$ and it is clear
from Table~\ref{table:Battack} that
\[
\tfrac{1}{\sqrt{2}}\I_B^\dagger\left(\ket{s_0} + \mathrm{i}\ket{d_0}\right) = \tfrac{1}{\sqrt{2}}(\ket{a_0} - \mathrm{i}\ket{b_0}) \in \tH_{\{0,1\}}.
\] 
Applying $\I_B$ to that states gives the state  $\frac{1}{\sqrt{2}}(\ket{s_0}+i\ket{d_0})$ which gives output $(z,0)$.

Similarly, in order to force a measurement of $(z,1)$,  using the same table we see that 
\[
-\J_B\ket{s_2} + \mathrm{i}\J_B\ket{d_2} = \ket{a_1} + \mathrm{i}\ket{b_1} \in \tH_{\{0,1\}}
\]
 and the  attack in Table~\ref{table:zoattack}
gives full information to Eve without affecting Bob's statistics given the current specification.

 \begin{table}[H]
\begin{center}
\renewcommand\arraystretch{1.25}
\begin{tabular}{ll}
\hline
\textit{\textbf{State Eve can send}} & \textit{\textbf{Bob's output}} \\ 
\hline
$(\ket{a_0} - \mathrm{i}\ket{b_0})/\sqrt{2}$ & $(z,0)$ \\
$(\ket{a_1} + \mathrm{i}\ket{b_1})/\sqrt{2}$ & $(z,1)$ \\
$(\ket{a_1} + \ket{a_{0}} - \mathrm{i}\ket{b_1} + \mathrm{i}\ket{b_{0}})/2$ $\phantom{\quad.}$ & $(x,0)$ \\
$(\ket{a_1} - \ket{a_{0}} - \mathrm{i}\ket{b_1} - \mathrm{i}\ket{b_{0}})/2$ & $(x,1)$ \\
\end{tabular}
\end{center}
\caption{A fixed-apparatus attack restricted to $\tH_{\{0,1\}}$}\label{table:zoattack}
\end{table}

Finally, recall that if there is no eavesdropping (no noise and no loss),  
if Alice sends~$\ket{0_z}$ and Bob gets output~$(z,0)$, 
then it is equally likely that he measured the state $\ket{s_0}$ or~$\ket{d_0}$.
Similarly, if Bob gets output~$(z,1)$, it is equally likely that he measured $\ket{s_2}$ or~$\ket{d_2}$. 
We note that also those statistics are preserved 
if Eve performs the attack of Table~\ref{table:zoattack} on~$\tH_{\{0,1\}}$.

\end{document}